\documentclass{ab}

\usepackage{graphicx}
\usepackage{epstopdf}
\usepackage{natbib}

\begin{document}

\title{Galactic Wind in NGC\,4460: New Observations}

\author{D.V.~Oparin$^{1*}$ and A.V.~Moiseev$^{2***}$}

\institute{$^1$Special Astrophysical Observatory, Russian Academy of Sciences, Nizhnij Arkhyz, 369167, Russia\\
$^2$Sternberg Astronomical Institute, M.V. Lomonosov Moscow State
University, Universitetsky pr., 13, Moscow, 119991
}

\titlerunning{GALACTIC WIND IN  NGC\,4460: NEW OBSERVATIONS}
\authorrunning{OPARIN, MOISEEV}

\date{May 19,  2015/Revised: July 31,   2015}
\offprints{$^*$D. Oparin  \email{doparin@mail.ru} $^{***}$ A. Moiseev \email{moisav@gmail.com}}

\abstract{
NGC\,4460 is an  isolated lenticular galaxy, in which galactic
wind has  been earlier discovered as a gas outflow associated with
circumnuclear regions of star formation. Using the results of
observations in the  H$\alpha$ line with the scanning Fabry--Perot
interferometer on the SAO~RAS 6-m telescope, we studied the
kinematics of the ionized gas in this galaxy. The parameters of
gas outflow from the  plane of the galactic disk were refined
within a simple geometric model. We show  that it is impossible to
characterize  the wind by a fixed velocity value. Characteristic
outflow velocities are within 30--80~km\,s$^{-1}$, and they are
insufficient to make the swept-out matter ultimately leave the
galaxy.
}

\maketitle

\begin{figure*}[]

\hspace{-3pt}
\includegraphics[scale=0.61]{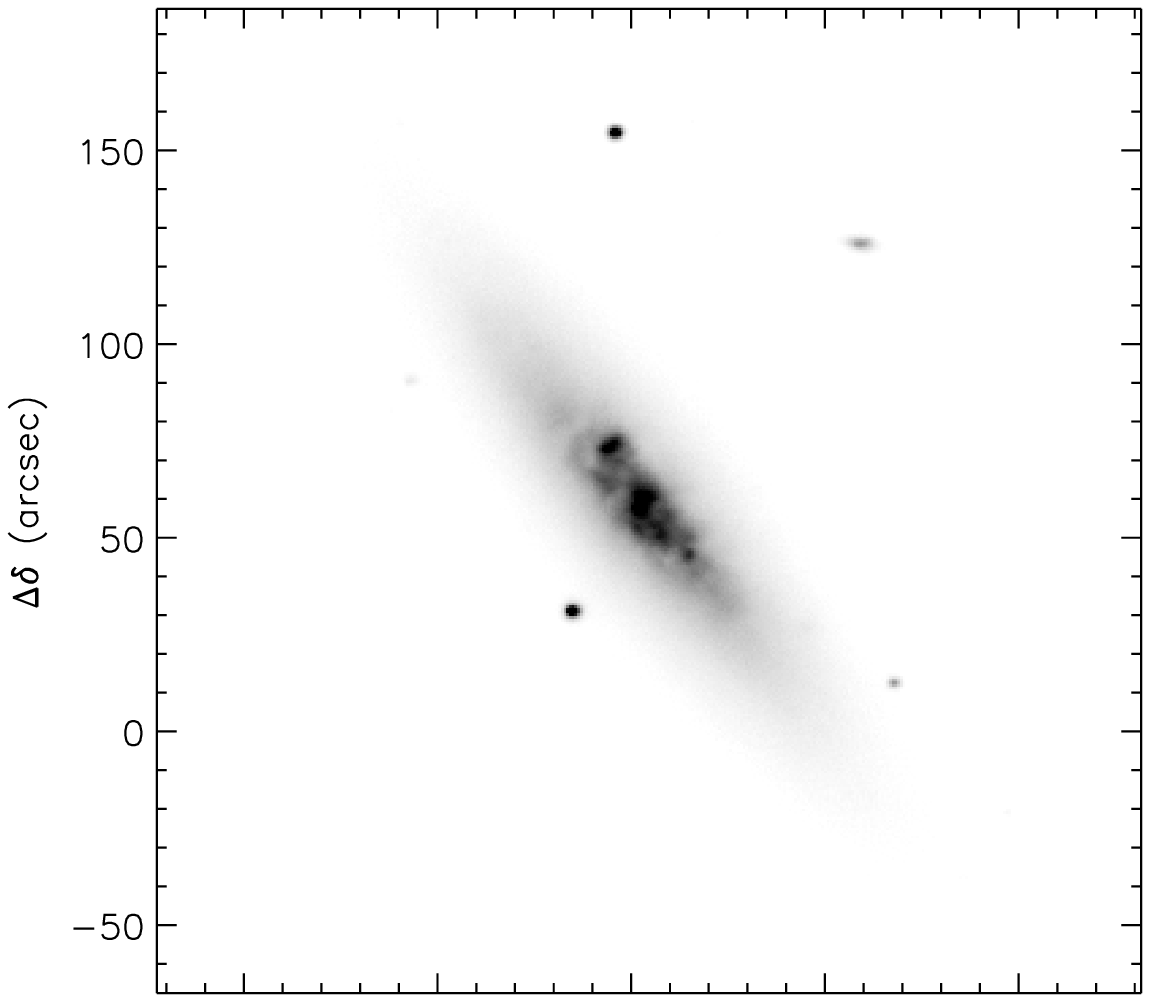}\hspace{5mm}
\includegraphics[scale=0.61]{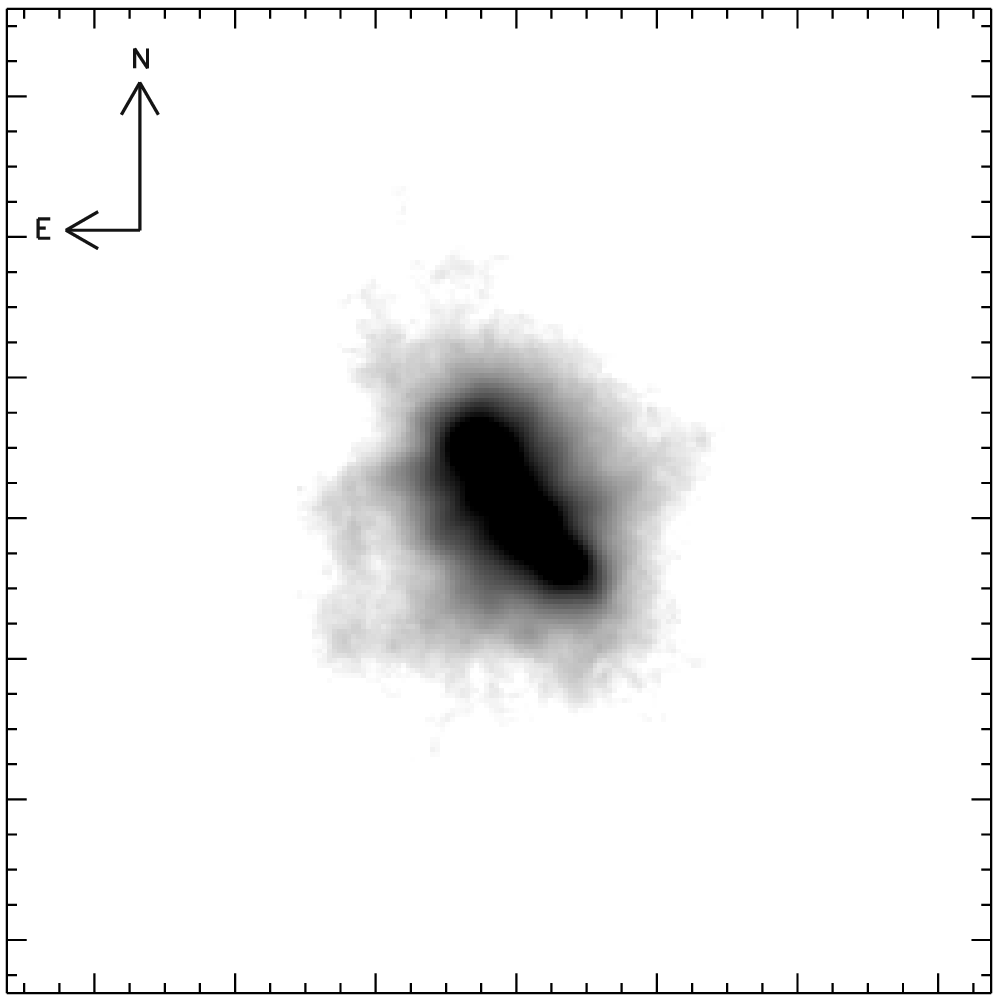}\\
 \vspace{5mm}
\includegraphics[scale=0.61]{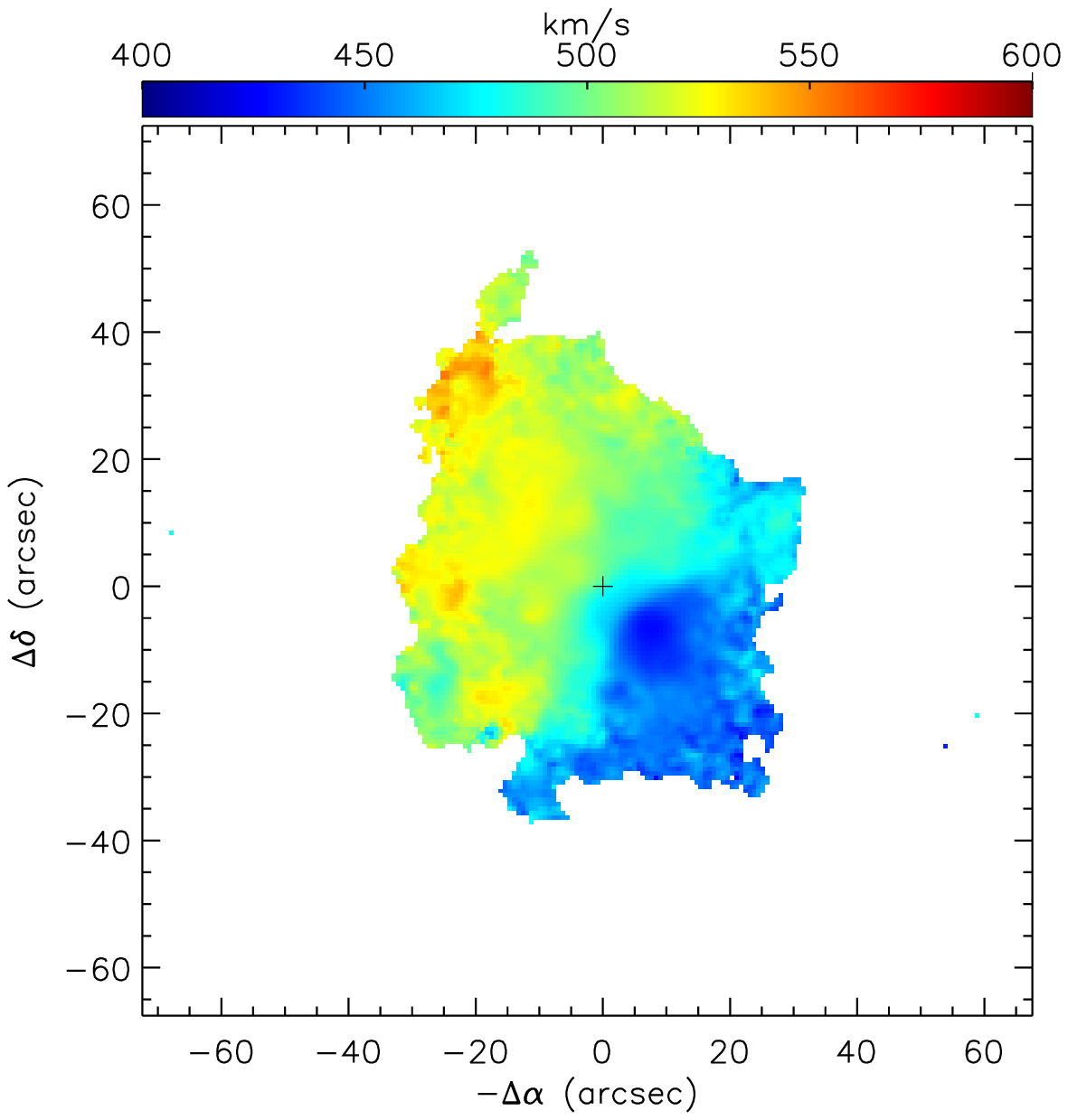}\hspace{5mm}
\includegraphics[scale=0.61]{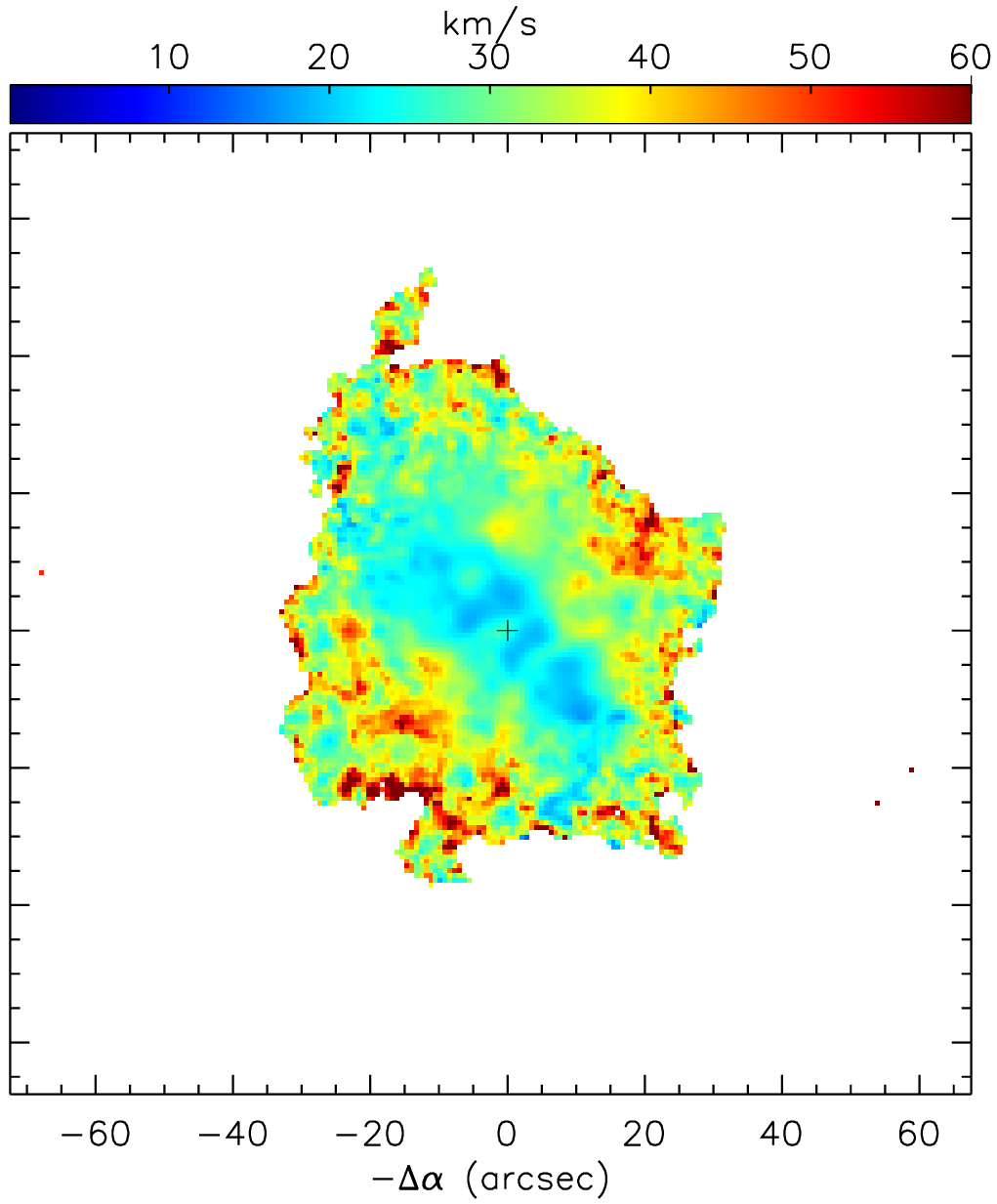}
 \vspace{-3mm}
\caption{Top left: an  SDSS image of the NGC 4460 galaxy in the $r$ filter; top right: an
image in the emission  H$\alpha$ line. Bottom left: the velocity
field; bottom right: the line-of-sight velocity dispersion field.}
\label{intro:Oparin_n}
\end{figure*}

\section{Introduction}

Galactic wind (galactic outflow, superwind, hereafter GW) is one
of the most impressive and large-scale manifestations of the
effect star formation has on the interstellar and even
intergalactic medium. It looks like a stream of gas (ionized,
neutral, molecular) emitted from a galaxy as a result of
collective action of the phenomena related to young stellar
groups: radiation pressure, winds of giant stars, supernova
explosions. GW has a serious impact on the structure and chemical
composition of the interstellar medium and mass distribution
therein,  stirring it, transferring the kinetic energy, and
starting the processes of star
formation~\citep[see, e.g.,][]{Veilleux2005:Oparin_n}. The results of numerical
calculations  underline the importance of the   GW role in
galactic evolution \citep{2012Hopkins:Oparin_n}, which
is also confirmed by the observations of  GW manifestations among
the galaxies at large redshifts~\citep{1998Pettini:Oparin_n}. The
observations of galaxies with intense star formation show that the
galactic outflow manifests itself in almost every one of them in
one way or another, at least as a broad pedestal at the base of
emission lines~\citep{2014Arribas:Oparin_n}.


Therefore, the GW phenomenon is well known 
\citep[see the review of][]{Veilleux2005:Oparin_n} and  references to
yet earlier studies therein). At the same time, a detailed study
of emissive nebulae, created by GW, with quite a high spatial
resolution was performed only for a few nearby galaxies, such as
M\,82~\citep{2009Westmoquette:Oparin_n},
NGC\,253~\citep{Matsubayashi2009:Oparin_n},
NGC\,1569~\citep{NGC1569:Oparin_n},
NGC\,3079~\citep{NGC3079Cecil:Oparin_n}. Hence, many questions,
particularly those related to the GW in dwarf galaxies, remain
open. Depending on the wind energy and the mass  of the galaxy
itself, matter can  be both swept out into the intergalactic
medium and, not possessing a sufficient initial velocity, get back
to the galactic disk. In which cases is one or the other scenario
realized? This requires both   detailed theoretical calculations
for specific objects \citep[see, e.g.,][]{2013M82:Oparin_n} and new
observational data.

The  NGC\,4460  galaxy considered in this paper is a good
``testing ground'' for this kind of research. This is a dwarf
lenticular galaxy with no noticeable companions,  located in the
scattered Canes Venatici  cloud. According to the database of the
Local Volume galaxies~\citep{2012Kaisina:Oparin_n}, its absolute
magnitude is  \mbox{$M_B=-17.73$,} the total mass
\mbox{$M=6.4\times10^{9}~M_\odot$}, and the accepted distance to
it amounts to 9.59~Mpc, which gives an apparent    scale of about
47~pc/arcsec.

An extended emission nebula was discovered here within the
H$\alpha$-survey of nearby galaxies~\citep{Kaisin2008:Oparin_n}.
Its spectroscopic study is presented
in~\citet{Moiseev2010_4460:Oparin_n}. It has been shown that most
of the emission line radiation originates from the compact (with a
diameter of about 1~kpc) region in the galactic disk center, which
is also confirmed by the optical (HST, SDSS) and ultraviolet
(GALEX) images. At the same time, diffuse radiation in the
H$\alpha$ line is distributed along the axis of rotation of the
galaxy on both sides of the nucleus over long distances (up to~1.5
kpc). Ionization of   gas in these areas is due to the combined
influence of photoionization of young hot stars and shock waves
related to current star formation. This way, according
to~\citet{Moiseev2010_4460:Oparin_n}, the extended nebula in
NGC\,4460 is generated by the galactic wind. The outflow velocity
estimates made in the same study amounted to more than
130~km\,s$^{-1}$, the kinetic energy of the wind was found to be
$5.8\times10^{52}$~erg, and the characteristic time of formation
of the observed structure amounted to \mbox{10--12~Myr.}

The data on   the motions of  ionized gas in the GW region was
provided  in~\citet{Moiseev2010_4460:Oparin_n} based on only two
spectral cross sections with a resolution corresponding to
\mbox{${\rm FWHM}\approx110$~km\,s$^{-1}$} in the H$\alpha$ line,
the central kiloparsec region was also observed with a panoramic
spectrograph with a resolution
 \mbox{${\rm FWHM}\approx160$~km\,s$^{-1}$}.
  That was enough to measure the ionization state of the gas.
However,  higher resolution in line-of-sight velocity for the
largest possible number of emitting regions is desirable for a
detailed study of kinematics of the galactic wind. This was done
in this paper based on the observations with a scanning
Fabry--Perot interferometer~(FPI).

\begin{figure*}
\includegraphics[scale=0.65]{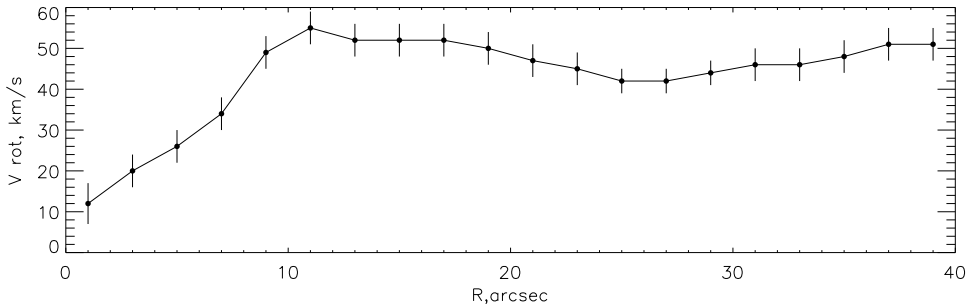}
 \vspace{-3mm}
\caption{Rotation curve of the galaxy.} \label{cir:Oparin_n}
\end{figure*}

\section{OBSERVATIONS AND DATA REDUCTION}

The observations were made on May~19 and 20, 2010 on the 6-m
telescope of the Special Astrophysical Observatory of the Russian
Academy of Sciences (SAO~RAS)  with a scanning FPI installed
inside the focal reducer
SCORPIO~\citep{AfanasievMoiseev2005:Oparin_n} in the H$\alpha$ line
with an average seeing of $2\farcs5$. The IFP751 interferometer at
the given wavelength provided a free spectral range
$\Delta\lambda=8.7$~\AA\ between the adjacent interference orders
and a spectral resolution of 0.4~\AA\ (19~km\,s$^{-1}$) at a scale
of 0.21~\AA\ per channel. The operating band around the redshifted
H$\alpha$ line was separated by a narrowband filter.

Scanning consisted of a series of 40 interferograms (two
$40\!\times\!80$-s series) obtained with different distances
between the interferometer plates, evenly filling the spectral
range. The result of data reduction, conducted with an application
package running in the IDL
environment~\citep{Moiseev2002ifp:Oparin_n,MoiseevEgorov2008:Oparin_n},
is a data cube, where each pixel contains a 40-channel spectrum.

As a result of  approximation of the  H$\alpha$ emission line
profiles  by the Voigt function, we built the images in the
H$\alpha$ line and in the continuum  as well as the line-of-sight
velocity and velocity dispersion maps (Fig.~\ref{intro:Oparin_n}).

\begin{figure*}[]
 \vspace{1mm}
\includegraphics[scale=0.65]{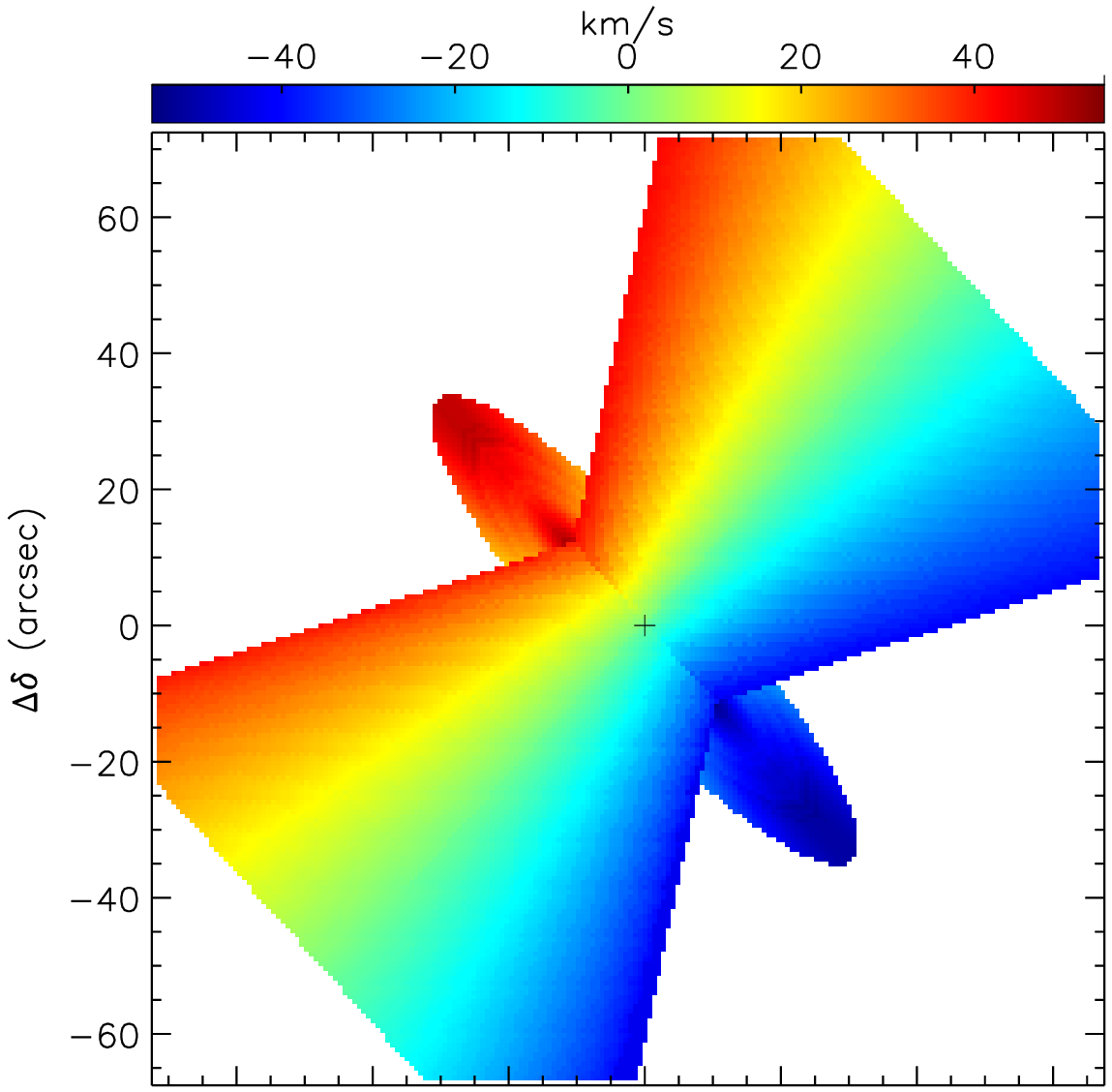}\hspace{5mm}
\includegraphics[scale=0.65]{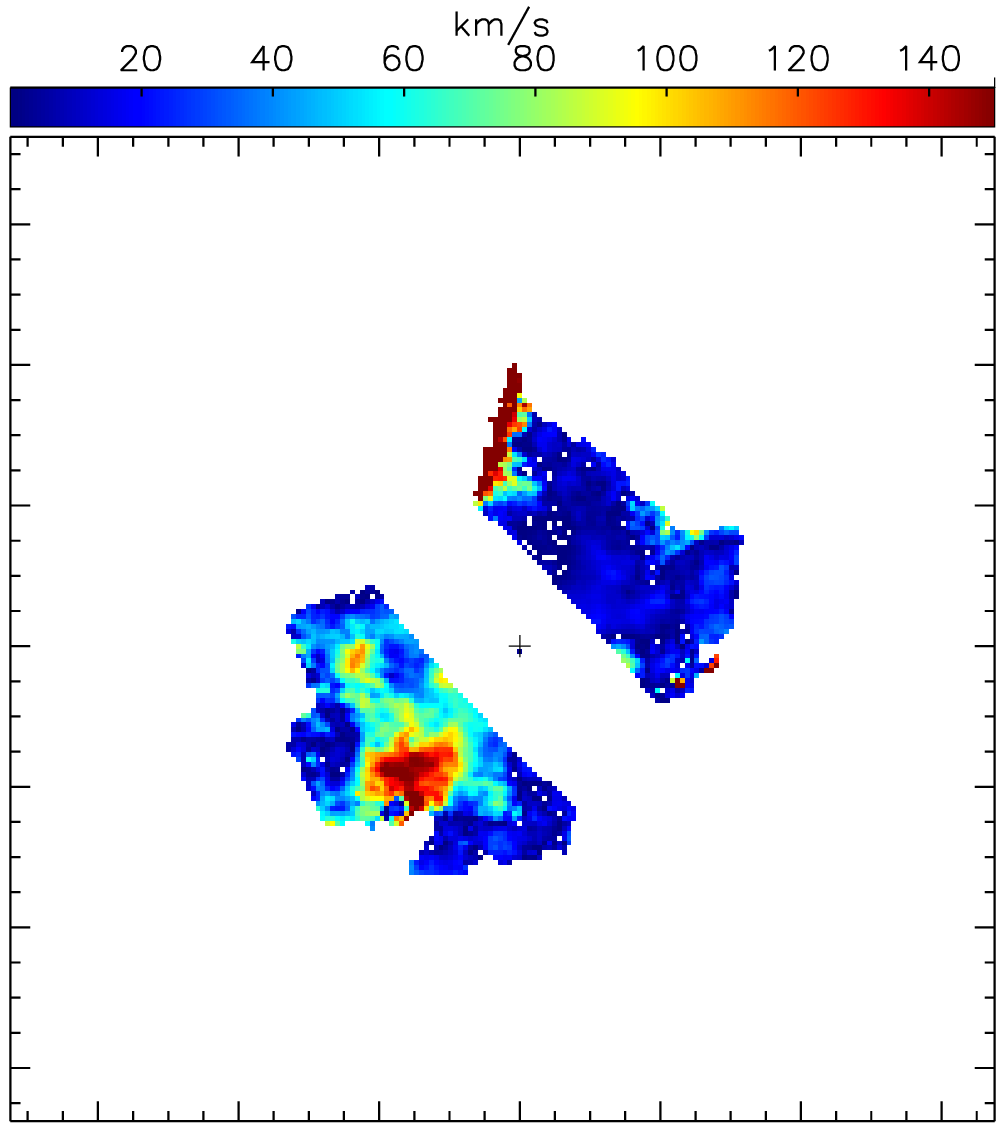}\\
 \vspace{5mm}
\includegraphics[scale=0.65]{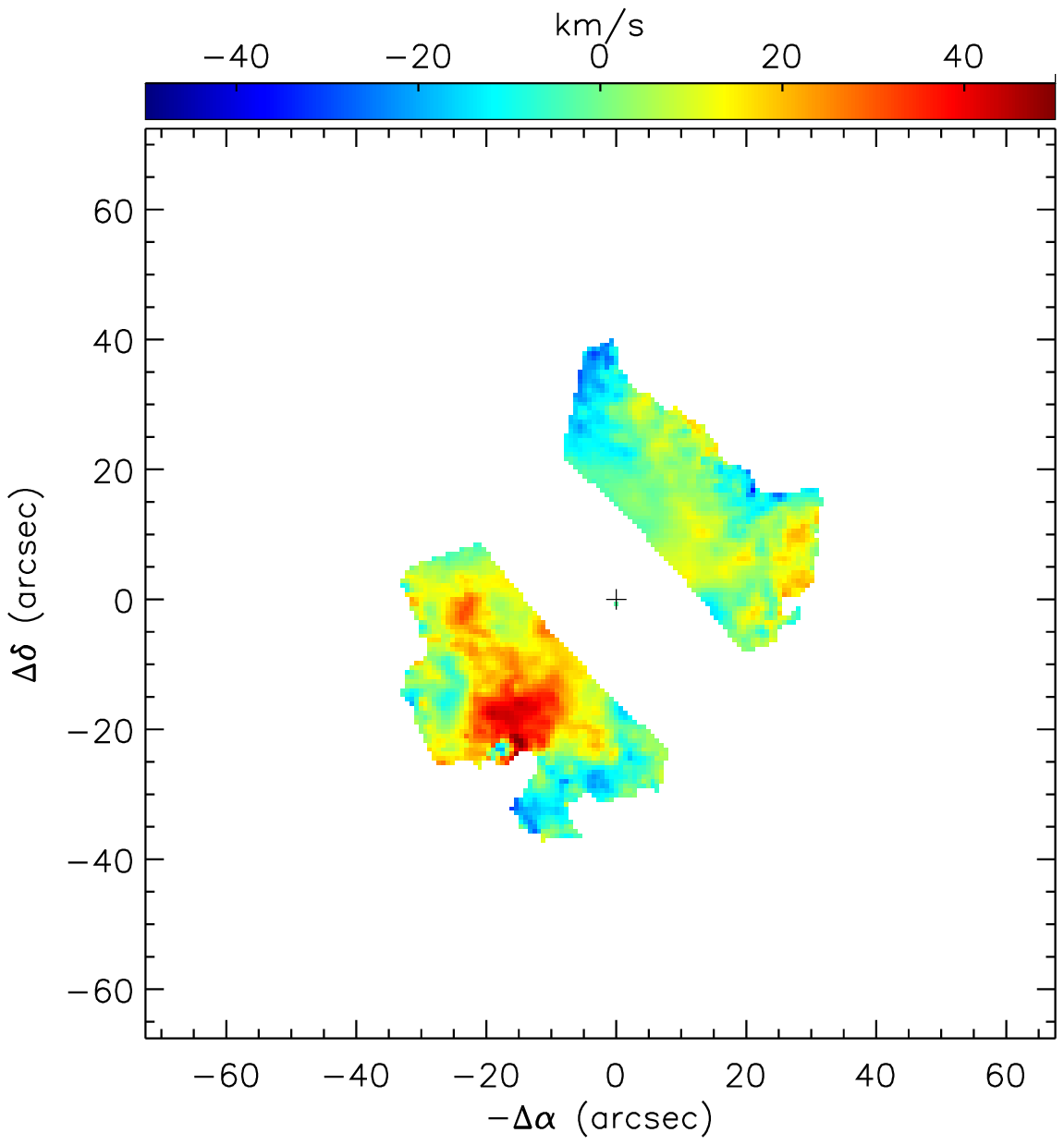}\hspace{5mm}
\includegraphics[scale=0.65]{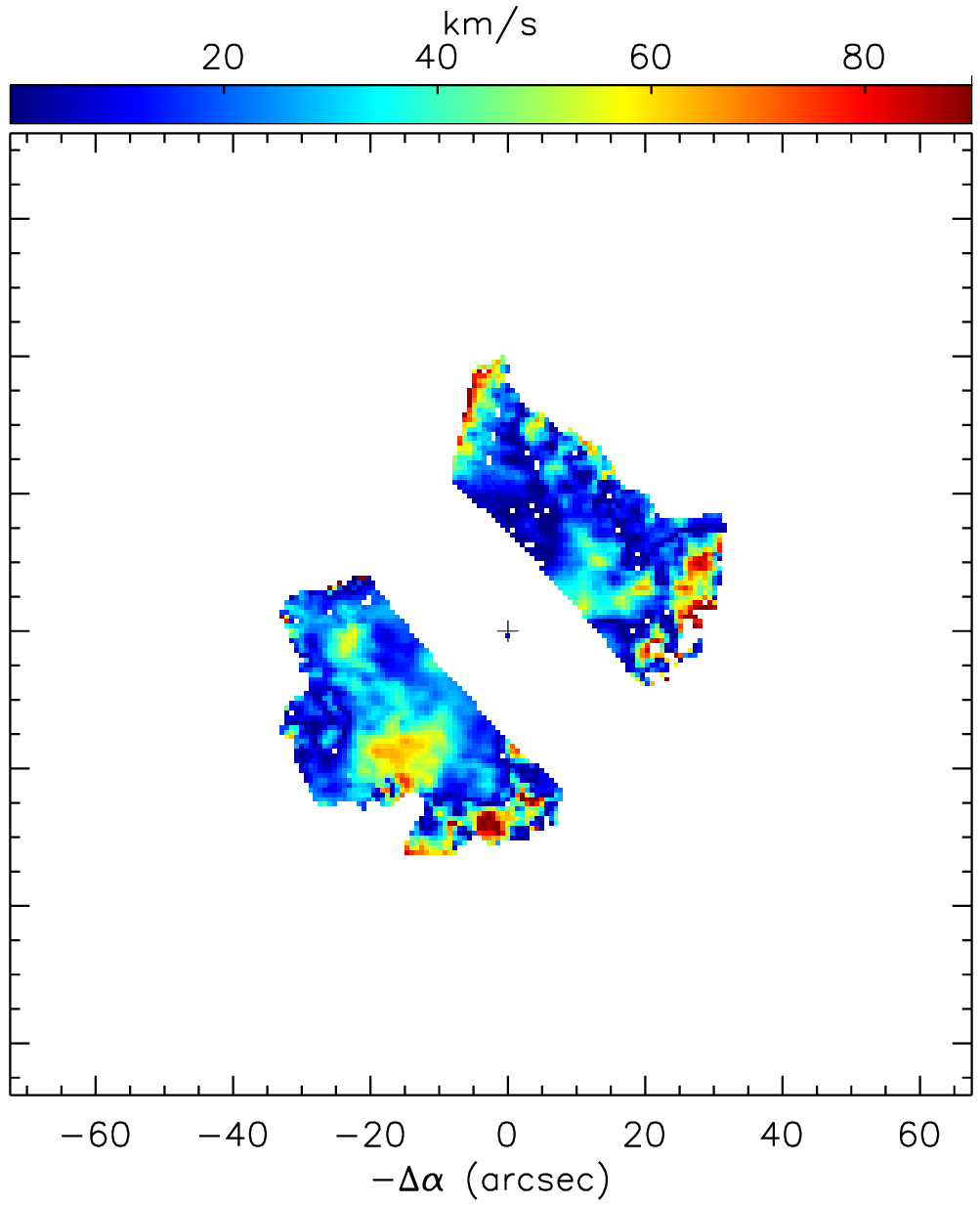}
 \vspace{-3mm}
\caption{Top left: a model  of rotation of the galactic disk and
swept-out matter of the galactic wind; bottom left: the map of
residual velocities, the central regions are masked. Top right:
the swept-out matter velocity map  along the walls of the cone
(the projection effect is taken into account) for the case when
the southeast cone is closer to the observer; bottom right: a
similar map for the case when the northwest cone is closer to the
observer. We believe the last orientation to be the most
plausible.} \label{residual:Oparin_n}
\end{figure*}

\section{ANALYSIS OF THE LINE-OF-SIGHT VELOCITY DISTRIBUTION}

\subsection{Galactic Disk Rotation Model}

The ionized gas velocity field derived from the  observations with
the FPI has already been used in~\citep{2014AstBu:Oparin_n} to
obtain an average rotation curve of the galaxy. Interestingly,
even outside the plane of the disk, in the GW region,  the
line-of-sight velocity distribution is dominated by the component
corresponding to circular rotation. Apart from the stellar wind,
the rotation of the galaxy itself also contributes to the
line-of-sight velocity of each point of the galaxy. Therefore, to
eliminate it, we had to create a model of circular rotation and
then subtract it from the velocity field. The rotation model was
built by the tilted-ring model under the assumption of a flat thin
disk (see~\citep{2014AstBu:Oparin_n} for details).

The following parameters, adopted
from~\citet{Moiseev2010_4460:Oparin_n}, were used to construct the
rotation model:
\begin{list}{}{
\setlength\leftmargin{2mm} \setlength\topsep{2mm}
\setlength\parsep{0mm} \setlength\itemsep{2mm} }
 \item ${\rm PA}_0=42\degr\pm2\degr$, the position angle of the line of nodes;
 \item $V_{\rm sys}=487\pm1$~km\,s$^{-1}$, the systemic velocity;
 \item $i=77\degr$, the angle between the sky plane and the plane
of the galactic disk.
\end{list}
The rotation curve of the galactic disk, obtained from the
velocity field is shown in Fig.~\ref{cir:Oparin_n}. In the
following section we extrapolate this   rotation  curve to the
region occupied by the galactic wind.

\subsection{Wind Cone Modeling}

The outflow from the region of active star formation takes place
in directions perpendicular to the plane of the galaxy, as the
matter density in this direction is substantially lower than in
the equatorial plane. Thus, there occurs an ``emersion'' of the
expanding swept-out shell into a less dense environment, as is the
case in the galaxies explored in detail, such as NGC\,1482,
NGC\,3079, or M\,82 (see the references in the Introduction). The
observed velocity field is dominated by circular rotation. This
indicates that the swept-out matter does not meet any serious
resistance. It can hence be assumed that its rotation moment  is
maintained. Based on the above, we can build a model of rotation
of the ejected shells.

We have assumed that the matter ejected  from the active star
formation region outflows as one large  structure---a shell. Such
a shell can be approximated by a frustum of a cone whose axis
passes through the galactic center and the smaller base is located
in the galactic disk. Here, only the walls of this cone could
basically be observed in the H$\alpha$ line, while the hot gas
inside does not radiate in the Balmer emission lines. It is worth
noting that these walls do not absolutely have to be continuous,
since in other galaxies with GW, the cones emitting in   optical
lines usually consist of separate filaments formed under the
effect of various kinds of
instabilities~\citep{Veilleux2005:Oparin_n}. Several such filaments
are also visible in the H$\alpha$ image of NGC\,4460.

The opening angle of the cone is assumed to be~$60\degr$, which
describes  the observed structure quite well.

When modeling the cone, we assumed that the outflowing matter is
not experiencing any significant  rotation velocity losses, since
the rotation moment is preserved  (see above), and the matter is
not ejected from the nucleus  itself but from an extended
kiloparsec area, and hence the radial motion  from the axis of
rotation is not substantial. Within this assumption, the
contribution of rotation in line-of-sight velocity is not
dependent on which side of the cone we observe, whether it is the
closest or the farthest to us, and which of the cones is closer to
us.

\begin{figure*}[]
\includegraphics[scale=0.66]{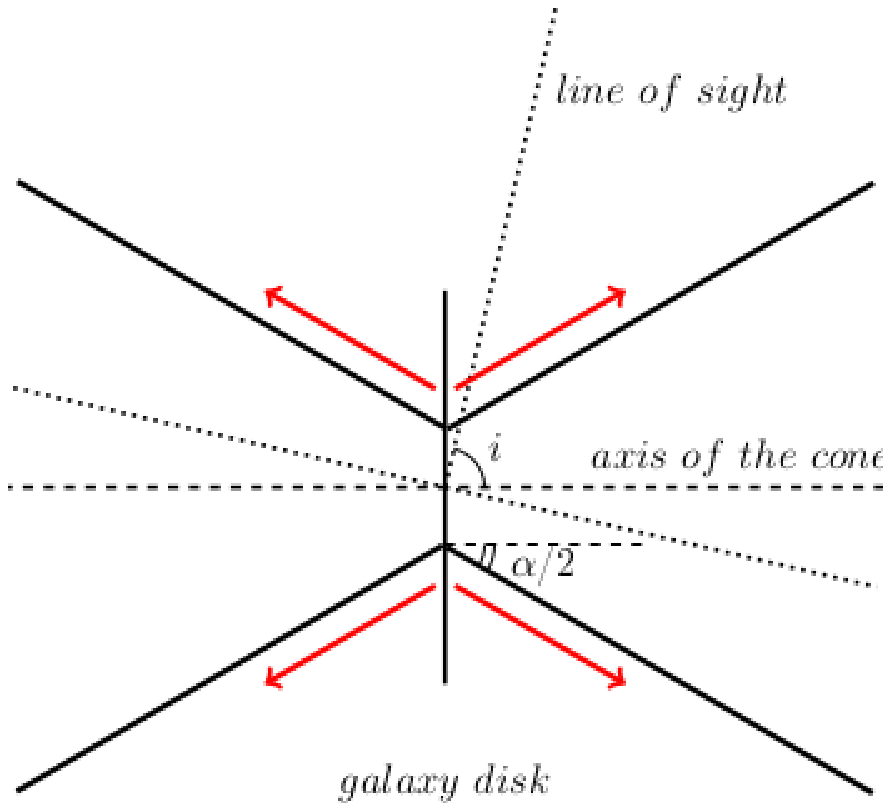}\hspace{20mm}
\includegraphics[scale=0.66]{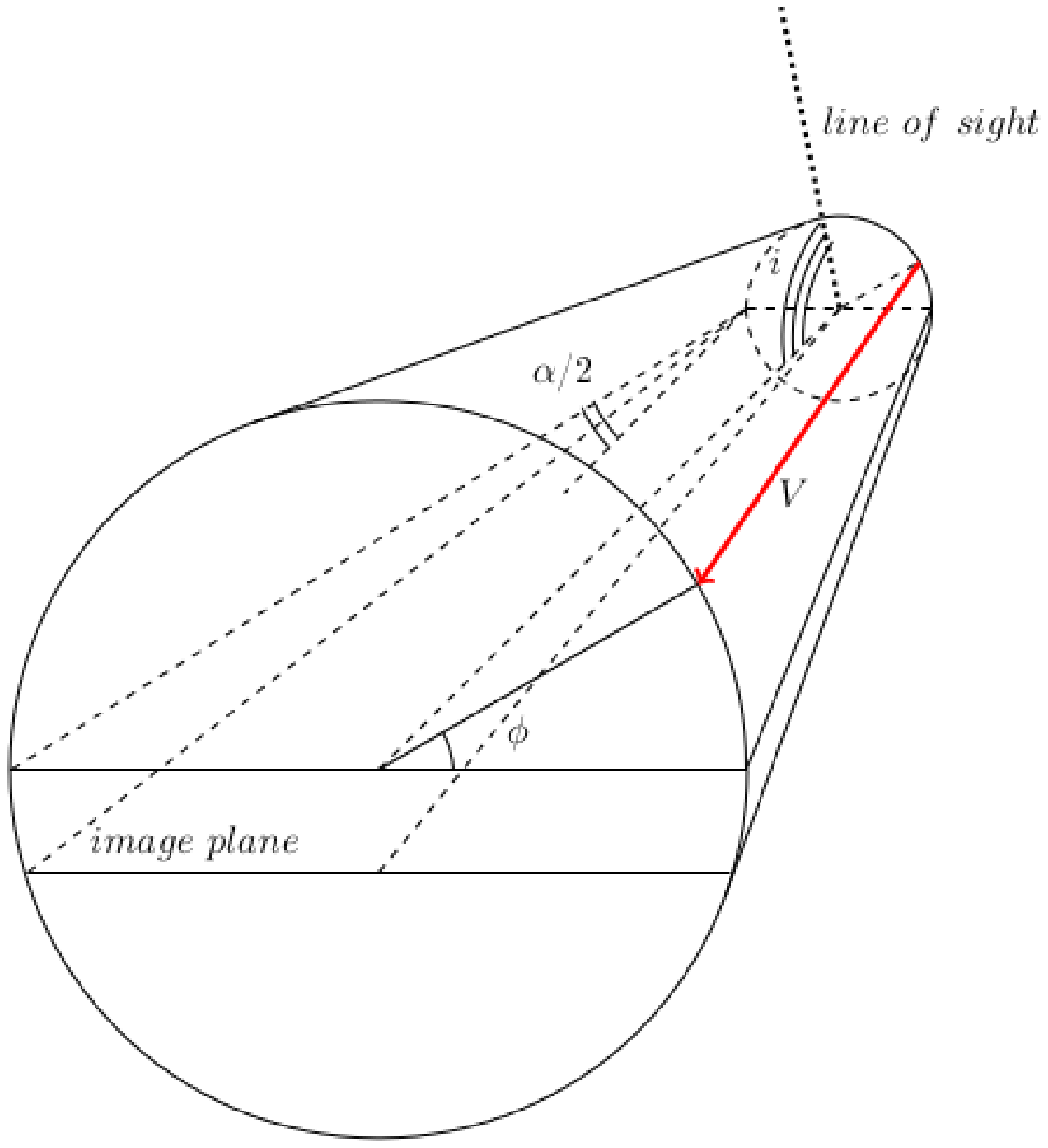}
\caption{Scheme of gas motion in the cone depending on the spatial
location. Left: the view from the plane perpendicular to the disk
of the galaxy; right: the view from the side of the cone axis. The
line of sight shows the direction to the observer. We can see that
if the  $\phi$ angle is sufficiently small, i.e., the point lies
close to the visible boundary of the cone, then for both its
possible positions (on the near and far walls of the cone from the
observer) the outflow velocity vectors will be directed to one
side from the sky plane.} \label{ris:Oparin_n}
\end{figure*}

Rotation velocities  of the cone are taken according to the
accepted rotation curve (Fig.~\ref{cir:Oparin_n}). The inclination
angle of the galaxy is $i=77\degr$; notice that it is difficult to
determine the disk orientation relative to the observer based on
direct images.

\subsection{Residual Velocities}
\label{sec33:Oparin_n}

Subtraction of the wind  rotation  model from  the   velocity
field produces  the map of residual velocities, shown in
Fig.~\ref{residual:Oparin_n} (bottom left). We can see no
significant velocity gradients in the direction of rotation,
which allows to consider our suggestion on   small differences
between the rotation velocities  in the disk  and wind to be
adequate enough.

Since the axis of rotation of the galaxy is inclined from the sky
plane by only $13\degr$, and assuming that the entire gas observed
in the cones  belongs to the walls and moves only in the direction
from the disk, we can assert that the negative residual
line-of-sight velocity supposes the motion on the wall of the cone
closest to us, while the positive velocity would imply the
farthest wall. The region where this rule would not work, as the
sections of both walls of the cone is on the same side of the sky
plane, is less than ten percent of the cone  radius at that point,
namely, within two to three pixels at the edges of the cone
(Fig.~\ref{ris:Oparin_n}). Thus, after subtracting the model of
rotation and systemic velocity, we can determine the spatial
distribution of gas emitting in  H$\alpha$ as well as on the
velocities of its rotation.

The calculation of   velocities is carried out according to the
formula derived from simple geometric considerations
(Fig.~\ref{ris:Oparin_n}):
$$ V_{\rm out}=\displaystyle{\frac{V_{\rm res}}{\cos(\alpha/2)\,\cos i+\sin(\alpha/2)\,\sin i\,\sin\phi}},$$
where $V_{\rm out}$ is the sought velocity of  gas moving along
the cone walls, $V_{\rm res}$ is the residual line-of-sight
velocity after subtracting the circular component, $i$ is the
angle between the sky plane and the plane of the galactic disk,
$\alpha/2$ is the half  opening  angle of the cone projection,
$\phi$ is the azimuth angle relative to the axis of the cone.

\begin{figure*}
\includegraphics[scale=0.6]{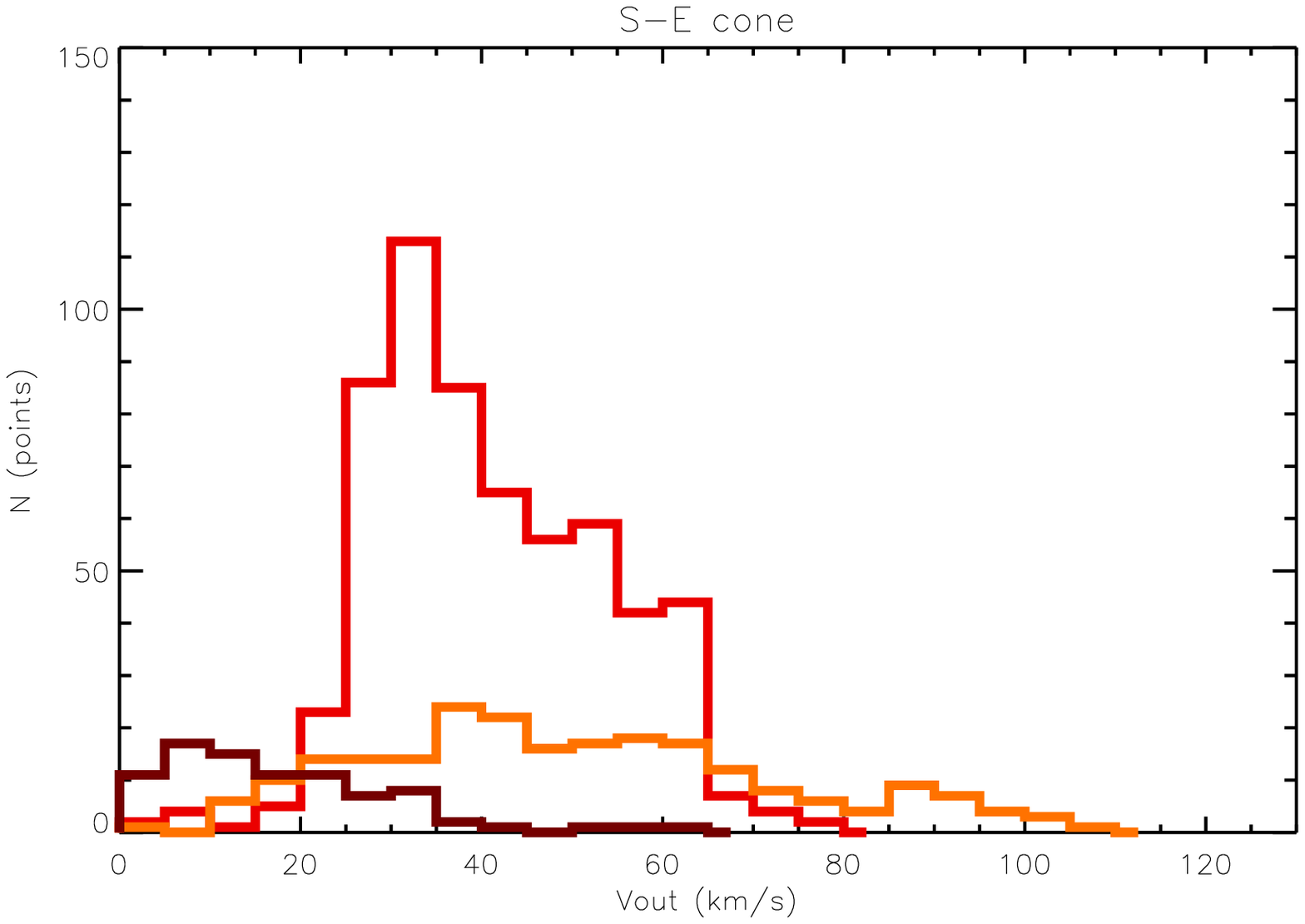}
\includegraphics[scale=0.6]{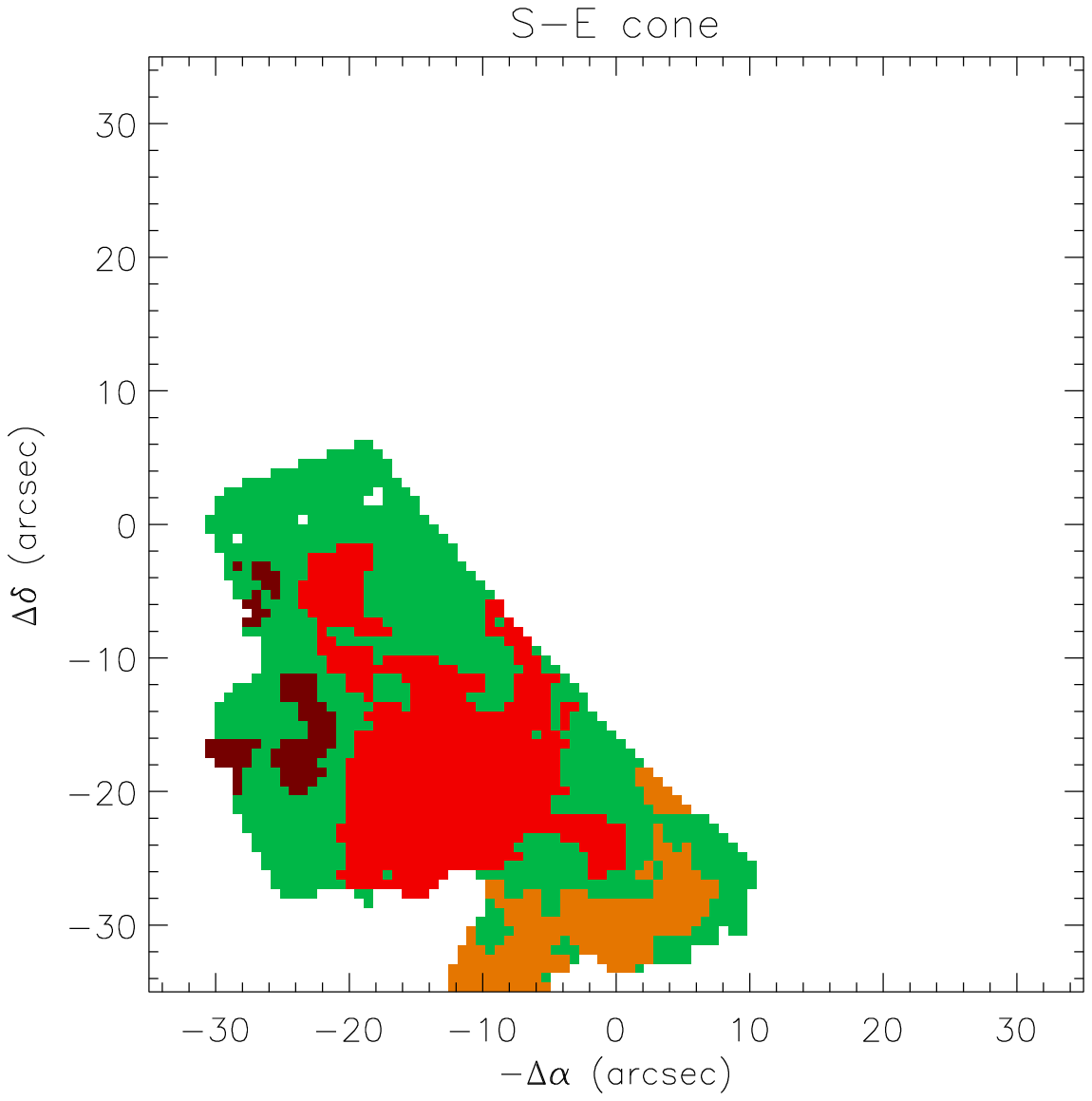}
\includegraphics[scale=0.6]{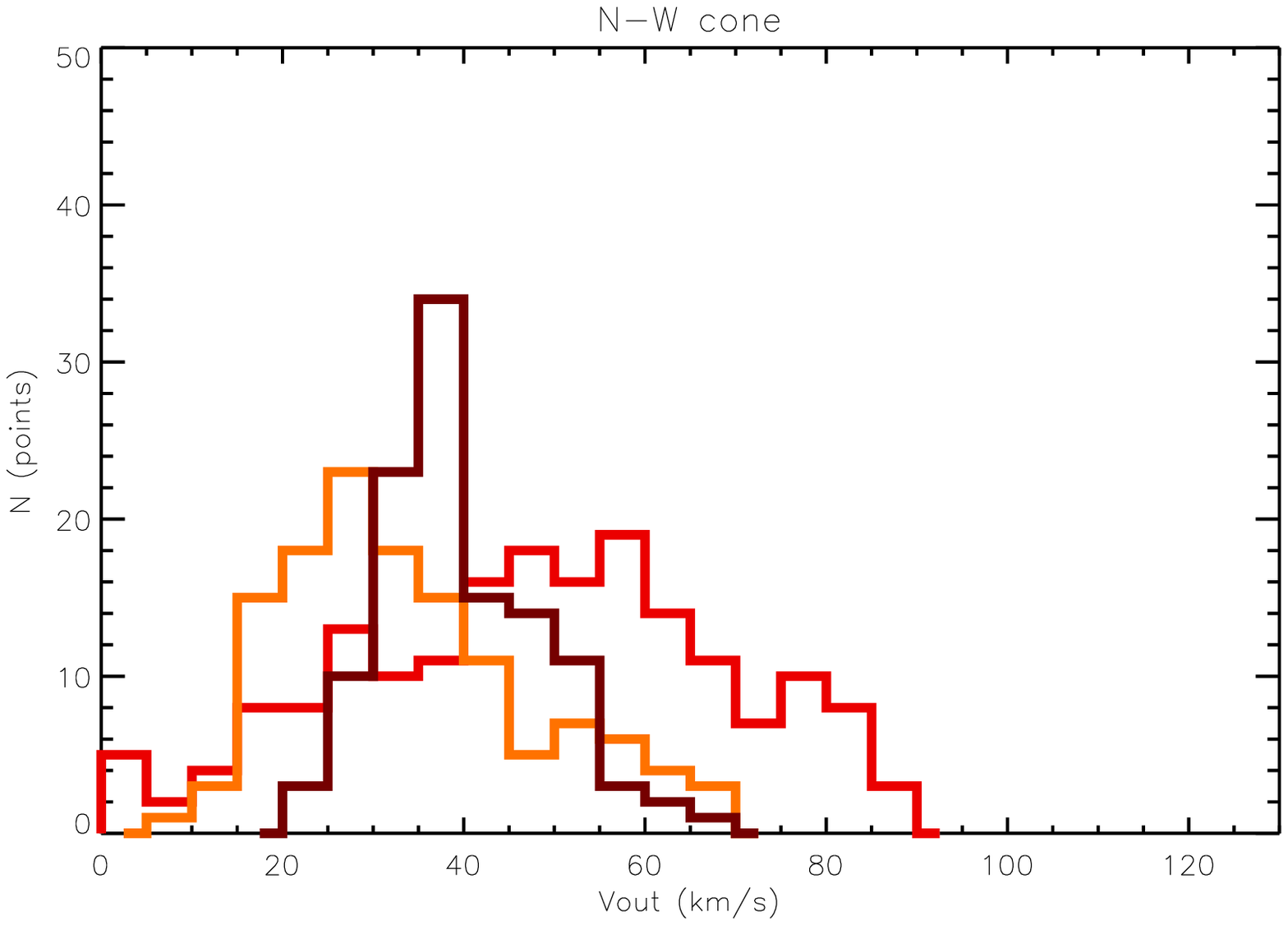}
\includegraphics[scale=0.6]{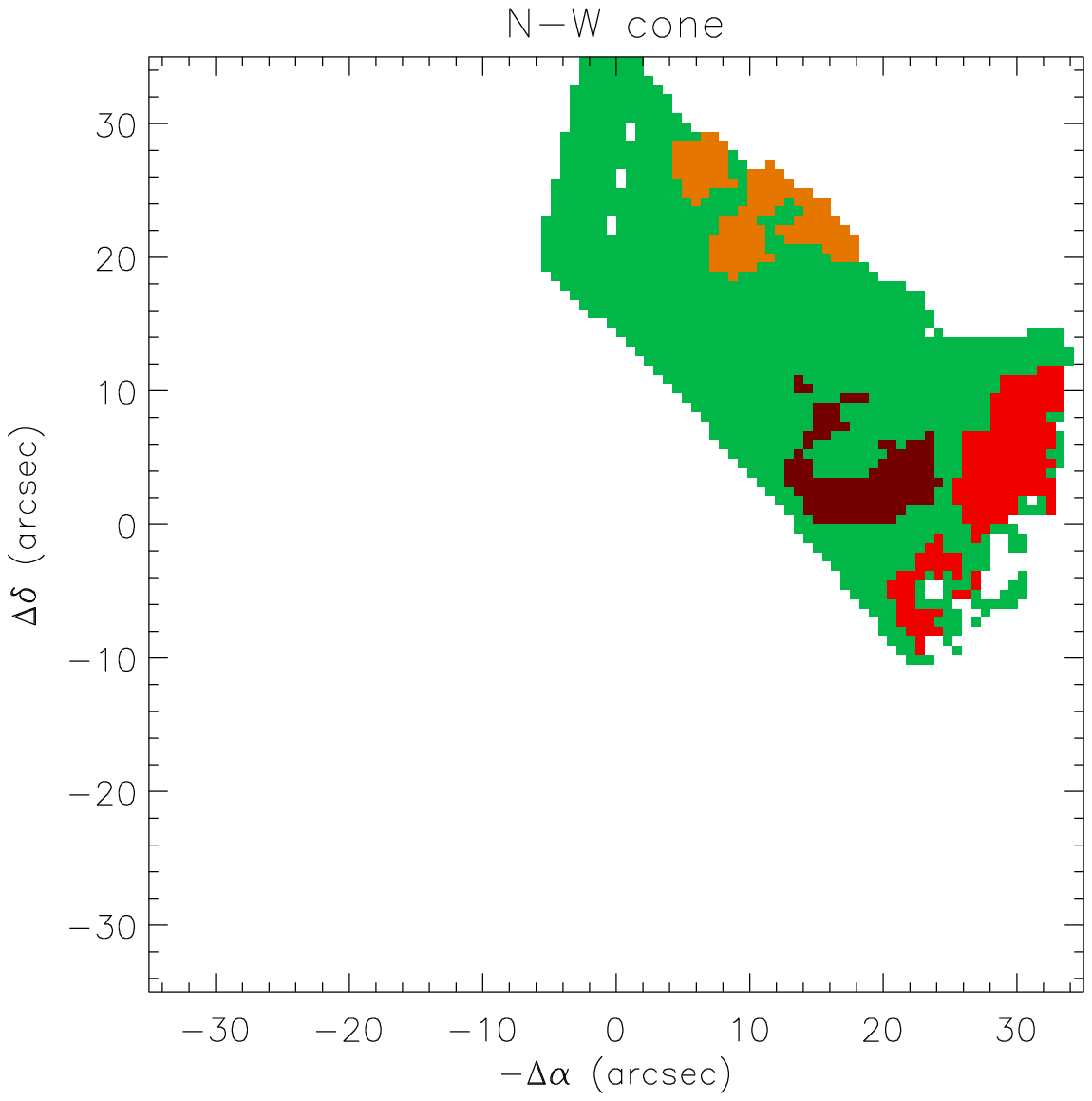}
\caption{Velocity distribution in the fastest filaments for the
far (southeastern, at the top) and near (northwestern, at the
bottom) cones. Right: locations of the  considered filaments on
the sky plane; left: the histogram of velocity distribution in
them. Green regions---background areas     with low
(\mbox{10--20}~km\,s$^{-1}$)  velocities--- are not represented in
the histogram. Colors  mark local areas with high velocities
of motion along the cone walls.} \label{hist:Oparin_n}
\end{figure*}

Because of uncertainty of the spatial orientation of the galactic
axis, it is not safe to say which of the model cones is closer to
the observer, which is required to recalculate the residual
line-of-sight velocities into the gas motion velocities along the
walls of the cone. We
 will hence consider both variants of cone orientation. Apart from that,
due to the  effect of projection on the picture  plane, there
occurs an uncertainty whether the radiating point is located in
the near or far wall of the cone relative to the observer. This
uncertainty affects the  $\phi$ angle. However, as we noted above,
knowing the sign of  residual line-of-sight velocities, we can
deduce on which side of the cone the emitting point is located.

Having  accounted for the projection effect,  we get  that the
model in which the southeast cone is accepted to be located closer
to the observer has a clearly different distribution of wind
velocities on both sides of the galaxy: 130--150~km\,s$^{-1}$ for
the southeast and 10--30~km\,s$^{-1}$ for the northwest cones;
while the second model (where the southeast cone is taken to be
the farthest from the observer) gives   much closer outflow
velocities for both cones. Therefore, further on we accept  this
spatial orientation. The high-velocity regions
at the cone boundaries are  of little use for the analysis, as
here we cannot clearly identify to which side, near or far, the
matter belongs, which means that we cannot calculate the  $\phi$
angle, and also due to the difference of our simple model from the
real morphology of the wind, which leads to obviously false
results at small $|\sin(\phi)|$  values.

\section{DISCUSSION}

 \subsection{Wind Parameters. Comparison with~Other~Galaxies}

The constructed   outflow velocity map
(Fig.~\ref{residual:Oparin_n}, bottom right) reveals individual
rapidly moving filaments on the background of relatively slow gas,
\linebreak \mbox{$V_{\rm out}=10$--$20$~km\,s$^{-1}$.}
Figure~\ref{hist:Oparin_n} gives the histograms of velocity
distribution within such regions.  The analysis of these
histograms allows us to conclude that the characteristic
velocities of the galactic wind in NGC\,4460 range within
30--80~km\,s$^{-1}$.

\begin{figure}[]
 \vspace{-1mm}
\includegraphics[scale=0.6]{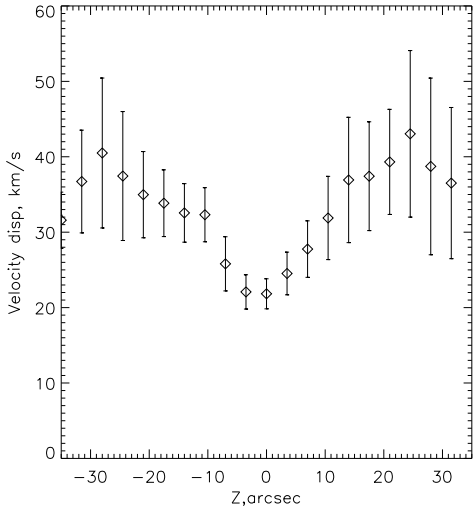}
\caption{Dependence of the velocity dispersion on the distance to
the galactic disk and the outflow velocity.} \label{disp:Oparin_n}
\end{figure}

When compared to other galaxies, such as\linebreak NGC\,1482 where
outflow velocities reach\linebreak
250~km\,s$^{-1}$~\citep{Veilleux1482:Oparin_n}, or
300--350~km\,s$^{-1}$ in M\,82~\citep{2009Westmoquette:Oparin_n},
the wind in NGC\,4460 shows significantly lower velocities,
comparable with the outflows in dwarf galaxies, such as NGC\,2366
\citep[\mbox{30--50~km\,s$^{-1}$--}][]{vanEymeren2008:Oparin_n} or
NGC\,4681
\citep[\mbox{30--80~km\,s$^{-1}$--}][]{vanEymeren2009:Oparin_n}.
Notice that in contrast to these two galaxies, where individual
isolated outflows from   H\,II regions were observed, a much more
large-scale and collimated structure of
 galactic wind is present in NGC\,4460.

Based on the line widths  measured in the low-resolution spectra,
a wind velocity   estimate   of more than 130~km\,s$^{-1}$  has
previously been obtained in~\citet{Moiseev2010_4460:Oparin_n}. Now
we have substantially refined this parameter, as our geometric
model allowed us to confidently separate the component
corresponding to the GW in the velocity field. This allowed us to
adjust the values of other wind parameters: the age of the
structure   formed by the wind lies within \mbox{20--50}~Myr; at
the mass of gas ejected from the disk \mbox{$M_{\rm
wind}=1.7\times10^5~M_\odot$},  its kinetic energy would be
 $E_{\rm wind}=0.3$--$2.2\times10^{52}$~erg.

The distribution of the line-of-sight velocity dispersion
perpendicularly to the galactic disk (Fig.~\ref{disp:Oparin_n})
shows that turbulent motions in  gas become substantial outside
the bright dense regions: velocity dispersion values there reach
40--50~km\,s$^{-1}$, which is comparable with the outflow velocity
itself, according to our estimates. Notice that a small
contribution to the measured velocity dispersion is brought in by
the thermal broadening of emission lines, it amounts to about
10~km\,s$^{-1}$ for an electron temperature of about 10\,000~K,
typical for the H\,II regions.

\subsection{The Galaxy Escape Velocity}

Using a procedure similar to that discussed\linebreak
by van Eymeren et al. (2009A,B),
we try to determine whether the wind velocity is sufficient to
overcome the galactic gravity and allow the gas to be ejected into
intergalactic space.

To estimate the velocity required for gas to leave the galaxy, let
us use the model of a spherical  pseudo-isothermal dark matter
halo, which describes well the outer curves of rotation for dwarf
galaxies.

In this model, the galaxy escape velocity is
$$ V_{\rm esc}(r)=\sqrt{2 V_c^2(1+\ln(r_{h}/r)},$$
where $V_c$ is the   rotation velocity of the galaxy,  $r_{h}$ is the
dark halo radius~\citep{Binney1987:Oparin_n}.

We assume the radius of the halo to be equal to the  Holmberg
radius of the galaxy of $2\farcm1$~\citep{2012Kaisina:Oparin_n},
which is about 6~kpc.

\begin{figure}[]
 \vspace{-1mm}
\includegraphics[scale=0.6]{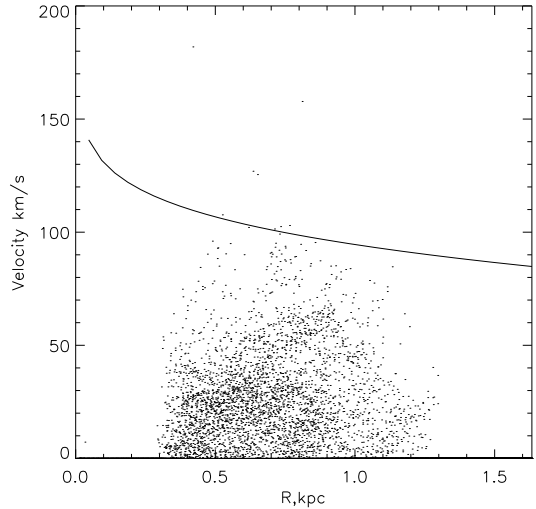}
\caption{Velocity  distribution of the matter outflow   from the
plane of the galactic disk (dots) depending on the distance to the
galactic center. The solid line represents the dependence of the
velocity required for the matter to escape from  the galaxy in the
model of a pseudo-isothermal halo with a radius of~6~kpc.}
\label{vesc:Oparin_n}
\end{figure}

The diagram of the velocity distribution of matter
outflow depending on the distance to the galactic center, is shown in
Fig.~\ref{vesc:Oparin_n}. If we impose on it the curve of the
model galaxy-escape velocity, we can see that the bulk of matter
is located considerably lower, i.e., it does not reach these
velocities. Several points formally located above the $V_{\rm
esc}(r)$ line are most likely due to the errors caused by a
different direction of the velocity vector than that expected in
Section~\ref{sec33:Oparin_n}.

Therefore, the outflow of  matter occurs with velocity insufficient
to escape the galaxy.

\section{CONCLUSION}

The observations with a scanning Fabry--Perot interferometer with
a sufficiently high spectral (about 19~km\,s$^{-1}$) and spatial
(about 120~pc) resolution allowed us to study the structure of the
galactic wind in NGC\,4460:
\begin{list}{}{
\setlength\leftmargin{4mm} \setlength\topsep{2mm}
\setlength\parsep{0mm} \setlength\itemsep{2mm} }
 \item[$\bullet$] the galactic wind in NGC\,4460 is mainly represented by two
streams outflowing from the galactic disk plane in opposite
directions; typical velocities of these streams are
\mbox{30--80~km\,s$^{-1}$;}
 \item[$\bullet$] the galactic wind in
NGC\,4460 is a ragged, very perturbed, and complex structure.
\end{list}

A comparison of  the velocities of  galactic wind  and the model
escape velocities for the case of pseudo-isothermal dark matter
halo suggests that the wind energy is not strong enough to eject
the gas beyond the galactic gravitational well. Therefore, we
expect  that over time, having cooled down, the matter will again
infall onto the galactic disk. This finding is consistent with the
similar results of van Eymeren et al. (2009A,B), obtained for  local outflows of ionized
gas in NGC\,2366 and NGC\,4681.

\begin{acknowledgements}
The work was supported by grant
\mbox{No.~14-22-00041} of the Russian Science Foundation. The
article is based on the data of observations carried out on the
SAO~RAS 6-m telescope with financial support by the Ministry of
Education and Science of the Russian Federation (agreement
No.~14.619.21.0004, project~ID PRFMEFI\,61914X0004).
\end{acknowledgements}

\end{document}